# An accelerated conjugate gradient algorithm to compute low-lying eigenvalues — a study for the Dirac operator in SU(2) lattice QCD[★]

Thomas Kalkreuter [1,2]

*Institut für Physik, Humboldt-Universität, Invalidenstraße 110, D-10099 Berlin, Germany*

and

Hubert Simma [1,3]

*Deutsches Elektronen-Synchrotron DESY, Platanenallee 6, D-15738 Zeuthen, Germany*

The low-lying eigenvalues of a (sparse) hermitian matrix can be computed with controlled numerical errors by a conjugate gradient (CG) method. This CG algorithm is accelerated by alternating it with exact diagonalizations in the subspace spanned by the numerically computed eigenvectors. We study this combined algorithm in case of the Dirac operator with (dynamical) Wilson fermions in four-dimensional SU(2) gauge fields. The algorithm is numerically very stable and can be parallelized in an efficient way. On lattices of sizes $4^4 - 16^4$ an acceleration of the pure CG method by a factor of $4 - 8$ is found.

---

[★] Work supported in part by Deutsche Forschungsgemeinschaft, grant Wo 389/3-1.
[1] Both authors are member of the ALPHA collaboration.
[2] Electronic address: kalkreut@linde.physik.hu-berlin.de
[3] Electronic address: simma@ifh.de

# 1 Introduction

Many problems in computational physics require the numerical determination of some of the low-lying eigenvalues of a (sparse) hermitian matrix $A$. For instance, $A$ may be a Hamiltonian which describes a many particle problem in quantum mechanics, or it may be the Dirac operator [4] in a lattice gauge theory. In case of a moderately large problem, the Lanczos method [1], or Cullum's and Willoughby's variant thereof [2], is popular. However, the Lanczos method can be problematic whenever one is interested in only a few eigenvalues.

In refs. [3–5] it was proposed to use a conjugate gradient (CG) method for the computation of low-lying eigenvalues. For the $k$-th eigenvalue one minimizes the Ritz functional [5]

$$\mu(z) = \frac{\langle z, Az \rangle}{\langle z, z \rangle} \ , \tag{1}$$

with $z \neq 0$ and orthogonal to the eigenspace of the $(k-1)$ lower eigenvalues. This method is attractive because it yields the eigenvalues with controlled numerical errors. A Lanczos method can be competitive in practice, but whether eigenvalues have converged can be estimated only from experience [6], and not from a rigorous error bound. Moreover, a Lanczos method cannot provide information about the multiplicities of eigenvalues, in contrast to the CG method. Some applications also require knowledge of the eigenvectors, for instance to isolate the contribution of low-lying eigenmodes to physical observables. For this purpose the CG algorithm is favourable because it also yields approximate eigenvectors.

In this article we investigate the CG algorithm in the version of ref. [5] and show that it can be accelerated by alternating incomplete CG minimizations with exact diagonalizations in the subspace spanned by the numerically computed approximate eigenvectors. We also improve the stopping criterion. This modified algorithm is studied in the context of lattice gauge theory where we take $A$ proportional to the square of $\gamma_5$ times the Dirac operator for massive Wilson fermions in four-dimensional SU(2) gauge fields. For an alternative way of combining CG searches with intermediate diagonalizations we refer to the work by Bunk [7].

Our interest in the low-lying eigenvalues of the lattice Dirac operator in QCD arises from their relation to chiral symmetry breaking [8] and from Lüscher's proposal [9,10] for the simulation of dynamical fermions. There, the small eigenvalues can be used to correct for possible systematic errors in case that the polynomial approximation to the function $1/s$ is too poor at the lower end of the spectrum.

---

[4] which is hermitean after multiplication with $\gamma_5$.
[5] $\langle \cdot, \cdot \rangle$ denotes the scalar product of the Hilbert space.

We expect that the results of the numerical studies presented in this paper translate to any operator which has a comparable spectrum to the (squared) Dirac operator. There, in a nontrivial background gauge field, the distribution of the eigenvalues is relatively smooth without exceptional gaps. On $4^4 - 16^4$ lattices we find an acceleration of the pure CG method by a factor of $4 - 8$, depending on the lattice size and the number of low-lying eigenvalues required.

This paper is organised as follows. In sec. 2 we briefly summarise the theoretical aspects of the CG algorithm of ref. [5], and we describe how it is accelerated and how the stopping criterion is improved. Practical aspects and the numerical implementation are discussed in sec. 3. In this section we also pay attention to questions of parallelisation of the algorithm, in particular on SIMD computers like APE/Quadrics systems. In sec. 4 we report performance tests of the algorithm in case of the lattice Dirac operator, and we draw conclusions in sec. 5.

## 2    Description of the algorithm

Before we describe our method in detail in the subsequent sections, we outline the idea behind the improvements of the CG method.

Variational methods to determine the lowest [6] eigenvalues of a hermitian operator $A$ in a Hilbert space require implicitly two steps: First, a suitable basis $\{w_1, w_2, \ldots, w_n\}$ has to be chosen for the subspace in which the eigenvectors are searched for (e.g. the 'trial wave-functions' for the ground state). Second, one has to determine linear combinations of these trial vectors such that the Ritz functional is minimised in appropriate subspaces of $\mathrm{span}\{w_1, w_2, \ldots, w_n\}$. The basic idea of our algorithm is to combine these two steps.

In fact, one can construct a reliable algorithm based only on step one by using the method of conjugate gradients [3–5]. Thereby, estimators $w_k$ of the correct eigenvectors $v_k$ of $A$ ($\|v_k\| = \|w_k\| = 1$) are constructed for each eigenvalue — one after the other and in increasing order, $k = 1, \ldots, n$. The expensive search for the different $w_k$ within the full high-dimensional space is thus done in an essentially independent manner, except for a projection on the subspace orthogonal to the $w_l$, $l < k$, which have already been found.

---

[6] We assume that $A$ is bounded from below. This will be the case in practical numerical work where $A$ can be represented by a finite matrix.

In order to accelerate this algorithm we alternate the CG searches for the $w_k$ with intermediate diagonalizations of the corresponding (small) hermitian matrix

$$M_{kl} = \langle w_k, A w_l \rangle \ . \tag{2}$$

Most of the eigenvalues of $M$ will be improved estimates for the low-lying eigenvalues of $A$.[7] For the eigenvectors improved estimates $w'_k$ are obtained through linear combinations of the $w_k$ with coefficients which correspond to eigenvectors of $M$. In the following the vectors $v_k$, $w_k$, and $w'_k$

After the diagonalization step the CG algorithm is restarted with initial vectors $w'_k$. Unfortunately, the efficiency of the CG method may suffer from this restart. This is, first of all, because a new system of search directions has to be built up, and second, because the intermediate diagonalization is unfavourable for the higher eigenvectors. Moreover, for the $(k+1)$-th eigenvector one is actually working with the matrix $Q_k^\perp A Q_k^\perp$ instead of $P_k^\perp A P_k^\perp$, where

$$Q_k^\perp = \mathbb{1} - \sum_{i=1}^{k} w_i \langle w_i, \cdot \rangle \tag{3}$$

can be a bad approximation to the projector

$$P_k^\perp = \mathbb{1} - \sum_{i=1}^{k} v_i \langle v_i, \cdot \rangle \ . \tag{4}$$

Therefore, the intermediate diagonalization may partly spoil the naively expected gain in convergence, and clearly, a careful balancing between CG search cycles and intermediate diagonalizations will be crucial in order to get an optimal trade-off.

Independent of the improvement of the rate of convergence, the intermediate diagonalization allows to speed up the algorithm by providing a better, i.e. more realistic, stopping criterion. It takes into account that the CG method converges proportional to the squared norm of the (complex) gradient of (1),

$$g(z) = [A - \mu(z)] z \ / \ \langle z, z \rangle \ , \tag{5}$$

rather than a linear behaviour, which can be concluded from the error estimate of refs. [4,5] but which is typically several orders of magnitude too pessimistic.

---

[7] There may also be some eigenvalue estimates which get worse.

## 2.1 Conjugate gradient algorithm to minimise the Ritz functional

In this subsection we recall the CG algorithm of ref. [5] for the computation of the lowest eigenvalue of a (sparse) hermitian matrix $A$. In contrast to the standard CG procedure for the minimisation of quadratic forms [1,11,12], positivity of $A$ is not required for the CG minimisation of the corresponding Ritz functional. We also note that due to the scale invariance of the Ritz functional (1) one has the orthogonality relation

$$\langle z, g(z) \rangle = 0 \quad \text{for any vector } z \neq 0. \tag{6}$$

The CG recursion starts with an arbitrary non-zero initial vector $x_1$ and a search direction $p_1 = g(x_1)$. Then, in the $i$-th iteration ($i = 1, 2, \ldots$) one computes the new vector

$$x_{i+1} = x_i + \alpha_i p_i \tag{7}$$

by choosing $\alpha_i$ in such a way that $\mu(x_{i+1})$ is minimised. Note that in the case of the Ritz functional this minimization can be carried out analytically. The new search direction is obtained by setting

$$p_{i+1} = g_{i+1} + \beta_i \left[ p_i - x_{i+1} \langle x_{i+1}, p_i \rangle / \langle x_{i+1}, x_{i+1} \rangle \right] \tag{8}$$

where $g_i = g(x_i)$, and $\beta_i$ is computed according to

$$\beta_i = \langle g_{i+1}, g_{i+1} \rangle / \langle g_i, g_i \rangle . \tag{9}$$

As in the case of the multidimensional minimisation of a general function with non-constant Hessian matrix, there is no unique criterion for the choice of the search directions $p_i$ and of the scale factors $\beta_i$ (for related works see refs. [3,4,7]). Eq. (8) ensures the orthogonality relation

$$\langle x_j, p_j \rangle = 0 \quad \text{for all } j \geq 1, \tag{10}$$

which is imposed here in accordance with the scale invariance of the Ritz functional (cf. eq. (6)), but which does in general not hold in standard CG algorithms [8]. Eq. (9) for the choice of $\beta_i$ is part of the definition of the algorithm and corresponds to the Fletcher-Reeves method [11,12].

---

[8] Usually the term proportional to $x_{i+1}$ in (8) is not present.

The orthogonality relation

$$\langle p_i, g_{i+1}\rangle = 0 \tag{11}$$

follows (as in the case of minimising a quadratic form) simply from the fact that $\mu_{i+1}$ is minimised along the line (7). Together with (8) and (6) this implies

$$\langle p_i, g_i\rangle = \langle g_i, g_i\rangle \ . \tag{12}$$

The algorithm terminates when $p_i = 0 \Leftrightarrow g_i = 0$, i.e. when $x_i$ is an exact eigenvector. Here "$\Leftarrow$" follows from (8), while "$\Rightarrow$" is a consequence of (12). In practice, one stops the recursion when $\|g_i\|$ is smaller than some threshold.

In ref. [5] it was shown that the algorithm is well-defined, i.e. the absolute minimum of $\mu(x_{i+1})$ along the line (7) is attained for a finite value of $\alpha_i$ unless $p_i = g_i = 0$ (and the algorithm terminates). Moreover, all $\alpha_i$ are real [5].

In a naive implementation of the above algorithm one might run into numerical difficulties because it follows from (7) and (10) that $\|x_i\|$ grows monotonically

$$\|x_{i+1}\|^2 = \|x_i\|^2 + \alpha_i^2 \|p_i\|^2 \ . \tag{13}$$

To circumvent this increase one can formulate the CG recursion entirely in terms of normalised vectors by making use of the scale invariance of the Ritz functional

$$x_i \to \frac{x_i}{\|x_i\|} \quad , \quad p_i \to p_i \cdot \|x_i\| \quad , \quad g_i \to g_i \cdot \|x_i\| \ . \tag{14}$$

A numerical implementation of the accordingly rescaled basic recursion can be set up in such a way that it may be considered as an operation working on "states" $(x, y, p, \mu, \|p\|, \|g\|)$ consisting of

- a unit vector $x$,
- the vector $y = Ax$,
- the current search direction $p$,
- the value $\mu = \mu(x)$ of the Ritz functional,
- the norm $\|p\|$ of $p$, and the norm $\|g\|$ of the gradient $g = Ax - \mu x$.

The initial vector $x$ may be chosen randomly, for instance, and the initial search direction $p$ is set equal to the gradient at $x$. The recursion then produces the next state $(x', y', p', \mu', \|p'\|, \|g'\|)$ from the current one in the following six steps.

(i) Check whether the stopping criterion is satisfied and exit if true.

(ii) Calculate $Ap$ and store the result in some auxiliary array $z$.
(iii) Compute[9] $\cos\theta > 0$ and $\sin\theta$ by minimising the Ritz functional along the circle $x\cos\theta + p/\|p\|\sin\theta$.
(iv) Compute $x'$ and $y'$ according to

$$x' = \cos\theta\, x + \sin\theta\, p/\|p\| \; , \tag{15}$$
$$y' = \cos\theta\, y + \sin\theta\, z/\|p\| \; . \tag{16}$$

(v) Compute $g' = y' - \mu' x'$ and store the result in the auxiliary array previously used for $z$.
(vi) Calculate the norm $\|g'\|$, the coefficient $\beta = \cos\theta\|g'\|^2/\|g\|^2$, and

$$p' = g' + \beta\,[p - x'\langle x', p\rangle] \; . \tag{17}$$

The algorithm is guaranteed to converge because the sequence of $\mu$'s is monotonically decreasing, and it is bound from below. In practice one has to halt the algorithm after a limited number of iterations. A safe stopping criterion could be based on the following rigorous error estimate which one verifies by expanding $x$ in an orthonormal basis of eigenvectors of $A$:

*If $x$ is a unit vector such that the gradient of the Ritz functional satisfies $\|g(x)\| < \omega$, then there exists an (exact) eigenvalue $\lambda$ of $A$ such that*

$$|\lambda - \mu(x)| < \omega \; . \tag{18}$$

## 2.2 Higher eigenvalues and intermediate diagonalization

In order to extend the CG method to the computation of further — degenerate or higher-lying — eigenvalues of $A$, the CG search directions are restricted to the subspace orthogonal to previously found eigenvectors. More precisely, if $v_1, \ldots, v_{n-1}$ denote the exact eigenvectors of $A$ corresponding to the $n-1$ lowest-lying eigenvalues $\lambda_1, \ldots, \lambda_{n-1}$, then an approximation to the next eigenvalue $\lambda_n$ can be determined by performing the CG minimisation of the functional $\langle z, P^\perp_{n-1} A P^\perp_{n-1} z\rangle/\langle z, P^\perp_{n-1} z\rangle$ where $P^\perp_{n-1}$ is given by eq. (4).

In practice, only approximations $w_1, \ldots, w_{n-1}$ to the exact eigenvectors are available, and the CG minimisation for the next eigenvalue can only be performed with $Q^\perp_{n-1}$ given by (3). Depending on the quality of the approximation of $P^\perp_{n-1}$ by $Q^\perp_{n-1}$, the vector $w_n$ resulting from the CG search, may then have a non-vanishing component in $\mathrm{span}\{v_1, \ldots, v_{n-1}\}$. This misorientation may be reduced by choosing a new basis $\{w'_1, \ldots, w'_n\}$ of $\mathrm{span}\{w_1, \ldots, w_n\}$ in which $Q_n A Q_n$ is diagonal, where $Q_n = \mathbb{1} - Q_n^\perp$. Such an intermediate diagonalization step yields also improved estimates for most of the eigenvalues of $A$ and rigorous error bounds analogous to eq. (18).

---

[9] In the algorithm where one works with normalised $x_i$, $\alpha_i$ is conveniently replaced by a real angle $\theta_i$ with $\cos\theta_i = \|x_i\|/\|x_{i+1}\|$, and $\sin\theta_i = \alpha_i\|p_i\|/\|x_{i+1}\|$ [5]. These quantities $\cos\theta_i$ and $\sin\theta_i$ as well as $\mu'$ can be computed purely algebraically.

The vectors $w_k'$ of the new basis are given by

$$w_k' = \sum_{l=1}^n \xi_l^{(k)} w_l \quad \text{for } k = 1, \ldots, n \tag{19}$$

where $\xi_l^{(k)}$ denotes the $l$-th component of the $k$-th normalized eigenvector of $M$, i.e. $\sum_{m=1}^n M_{lm} \xi_m^{(k)} = \nu_k \xi_l^{(k)}$ with $\|\xi^{(k)}\| = 1$. The linear combinations (19) are exact eigenvectors of $A$ restricted to the subspace spanned by the numerically computed $w_k$,

$$Q_n A Q_n w_k' = \nu_k w_k' , \tag{20}$$

and the eigenvalues $\nu_k$ of $M$ equal the corresponding values of the Ritz functional $\mu(w_k')$. The relation of these estimators to exact (possibly degenerate) eigenvalues of $A$ is clarified by the following lemma [5]:

*Suppose the gradient of the Ritz functional satisfies $\|g(w_k)\| < \omega$ for every $w_k$ in a set of orthonormal vectors $\{w_1, \ldots, w_n\}$ and let $\nu_k$ denote the eigenvalues of the matrix with elements $M_{kl} = \langle w_k, A w_l \rangle$. Then there exist $n$ orthonormal eigenvectors of $A$ with eigenvalues $\lambda_1, \ldots, \lambda_n$ such that*

$$|\lambda_k - \nu_k| < \omega \sqrt{n} \quad \text{for all } k = 1, \ldots, n. \tag{21}$$

This lemma shows that the eigenvalues are obtained with the correct multiplicities, which cannot be concluded from (18) applied to all the $\mu(w_k)$ separately.

The linear combinations $w_k'$, which again form an orthonormal set, can be expected to be better approximations to the true eigenvectors $v_k$ than the $w_k$. Using these rotated vectors $w_k'$ as the starting vectors for a subsequent cycle of CG searches leads us to the following algorithm:

(i) For each $k = 1, \ldots, n$ in succession, compute approximations $w_k$ to the eigenvectors $v_k$ of $A$ by performing only a certain number $N(k)$ of CG iterations for the minimisation of $\langle z, Q_{k-1}^\perp A Q_{k-1}^\perp z \rangle / \langle Q_{k-1}^\perp z, Q_{k-1}^\perp z \rangle$.
(ii) Compute the matrix $M$, diagonalize it, and determine the linear combinations (19). If necessary, reshuffle the indices $k$ in such a way that $\nu_1 \leq \ldots \leq \nu_n$.
(iii) Exit if a stopping criterion (see below) is fulfilled for all $k = 1, \ldots, n$. Otherwise, continue with the CG cycles in (i) using $w_k'$ as initial vectors (and possibly different search lengths $N(k)$).

This method of alternating CG cycles with intermediate diagonalizations of $A$ in the basis of approximate eigenvectors leads to a substantial acceleration of the algorithm, both in terms of the total number of CG iterations and in terms of computer time. An optimal choice of the numbers $N(k)$ for each cycle of CG iterations turns out to be a crucial ingredient. Our criterion for choosing $N(k)$ in an adaptive and automatic way will be discussed in sec. 3.3.

We note that the diagonalization step has an ambivalent character. Since $\sum_{i=1}^{n} \mu(w_i) = \sum_{i=1}^{n} \nu_i$ remains invariant during step (ii), it follows that whenever there are estimates $\nu_k < \mu(w_k)$, which are improved by step (ii), there are also — possibly worse — estimates $\nu_k > \mu(w_k)$. It turns out that typically just the last few $\nu_k$, with $k$ not much smaller than $n$, are lifted above the corresponding $\mu(w_k)$. This suggests to introduce a few "dummy" eigenvalues, i.e. if one is interested in the $n$ lowest eigenvalues, one runs the algorithm with $n + l$ eigenvalues ($l$ being a small integer number), but one does not demand the stopping criterion for $\lambda_k$ with $k > n$. Of course, if then all eigenvalues but the "dummies" have converged, one should not include the latter in the last diagonalization any more. Otherwise their lower precision could increase the error of the other eigenvalues, in particular the highest ones.

## 2.3  Error estimates based on Temple's inequality

Terminating the iterations only when $\|g(w_k)\| < \omega$ is fulfilled for all $k$, one ensures the rigorous error bounds (18) and (21). However, in realistic cases the actual error is found to be significantly smaller because the method does not only converge linearly (proportional to $\|g\|$) but rather quadratically (proportional to $\|g\|^2$). The latter is taken into account by an error estimate based on Temple's theorem [13,14].

Let $\lambda_1$ be the lowest (possibly degenerate) eigenvalue of a hermitian operator $A$. Provided $l_{>1}$ is a lower bound for the next higher eigenvalue denoted by $\lambda_{>1}$ with $\lambda_{>1} \geq l_{>1} > \mu(z)$, Temple's inequality yields a lower bound for $\lambda_1$ by

$$\mu(z) - \delta_T(z) \leq \lambda_1 \leq \mu(z) \quad \text{for all } z \neq 0, \tag{22}$$

where

$$\delta_T(z) = \frac{\frac{(z, A^2 z)}{(z,z)} - \mu(z)^2}{l_{>1} - \mu(z)} . \tag{23}$$

Rewriting the numerator in (23) in terms of $\|g(z)\|^2$ we obtain the error esti-

mate

$$0 \leq \mu(z) - \lambda_1 \leq \frac{\langle z, z \rangle}{l_{>1} - \mu(z)} \, \|g(z)\|^2 \; , \tag{24}$$

which shows that the error is bound by $\|g\|^2$.

Generalizations analogous to (24) hold for higher eigenvalues as well [13,14] and can be used as a more efficient stopping criterion than the bound from $\|g\|$. In the numerical studies carried out, we found that the true error is described well-enough for the purpose of error estimation by a constant times $\|g\|^2$. This behaviour will also be exploited by a third stopping criterion which we discuss in sec. 3.4.

## 3 Numerical implementation of the algorithm

In this section we shall present some details of the numerical implementation of the algorithm described in the previous section. Because we have in mind numerically extensive applications we will stress aspects of parallelisation. Since we have performed our studies on APE/Quadrics parallel computers, we shall also pay attention to questions of the parallelisation on single-instruction-multiple-data (SIMD) architectures and of the numerical stability in view of restricted 32-bit floating point arithmetics.

### 3.1  CG minimisation of the Ritz functional

The simplest way to parallelise the CG part of the algorithm is by geometrical data decomposition, where the "large" vectors $x$, $p$, $y$ and $z$ are partitioned into sub-vectors of equal length, each of which is treated (and stored) on a different processor node. The main part of floating-point operations to be performed on the vectors, like scalar products, linear combinations or the multiplication with the matrix $A$, can then be done simultaneously on all nodes for the locally stored sub-vectors. Assuming that the matrix $A$ is sparse and local, like the Wilson-Dirac operator considered below or any other similarly discretized differential operator, only a few nearest neighbour communications are needed for the matrix-vector multiplication provided that the data partitioning respects the geometrical structure of the original physical system which is investigated.

Scalar products require a simultaneous local summation over the components of the sub-vectors on each node followed by a global summation over all nodes to collect the partial results from the products of the sub-vectors. In order to avoid accumulation of rounding errors in these summations, which may involve a large number of terms, we use Kahan's formula for the local summation and an effective double precision method for the global summation [15].

The program can be organised in such a way that also multiple replica of a physical system, i.e. different matrices $A$, can be treated in parallel on subsets of the nodes of the whole machine. Of course, on a SIMD architecture one then introduces the problem that some of the processors may become idle if not all systems have converged at the same time. This cause of inefficiency is significantly reduced in our algorithm as a result of the intermediate diagonalizations (see sec. 4.2).

The CG recursion itself is stabilized by renormalizing the state (cf. sec. 2.1) every 10–50 CG iterations (and at the end of each CG cycle). This amounts to readjusting the length of $x$ to unity, recalculating $y = Ax$, and subtracting a linear combination of $x$ and $g$ from $p$ to ensure the relations eqs. (10) and (12). Moreover, the norms of $p$ and $g$ and the current value of the Ritz functional are recalculated. As further safety measures we use numerically stable formulas in the minimisation of the Ritz functional on the circles (15), and the magnitude of $\beta$ in eq. (17) is limited in order to avoid that the search vectors are becoming large. (If this cutoff comes to effect it just amounts to partially restarting the CG algorithm). In the case of higher eigenvalues $k > 1$ the renormalization of the state also includes the projection of $x$ and $p$ on the orthogonal complement of the previously computed approximate eigenvectors.

In order to ensure that $x$ and $p$ remain orthogonal to the previously determined eigenvector approximations, the corresponding projection with $Q_k^\perp$ is in principle required once per CG iteration when evaluating the matrix-vector product $z = Ap$. Since the number of vector operations[10] involved in $Q_k^\perp$ grows proportional to $k$, a significant or even dominant fraction of the CPU-time is spent on these projections already for moderately large $k$. Fortunately, one does not observe any instabilities or significant effects when skipping these projections. It is sufficient in practice to perform them only together with the state renormalizations mentioned above. This indicates that already after a few iterations, the approximate eigenvectors are sufficiently well oriented into the direction of the exact eigenvectors and that the algorithm is numerically very stable.

---

[10] Scalar products and linear combinations which themselves require an absolute CPU-time roughly proportional to the linear dimension of $A$

## 3.2 Jacobi diagonalizer for hermitian matrices

For the diagonalization of the small matrix $M$ of eq. (2) we use the well-known method of iterative Jacobi rotations [16,17,1,12] which is conceptually very simple and numerically foolproof. Other methods, such as repeated Householder reflections followed by diagonalization of the resulting tridiagonal matrix (e.g. by the QL or QR algorithm or by bisection), are believed to be superior for larger matrices, at least on serial or vector computers [17,1,12]. On the other hand, the Jacobi algorithm is straightforward to implement on a SIMD machine because of its simple global flow (no pivoting as might occur in case of a Householder transformation) and it can exploit the situation when $M$ is almost diagonal. Moreover, the Jacobi method behaves well in cases when there are multiple or close eigenvalues [18], and it converges quadratically.

By means of unitary similarity transformations the hermitian matrix $M$ is transformed iteratively into diagonal form without creating an intermediate tridiagonal matrix. In each complete Jacobi sweep one visits all off-diagonal elements in a fixed order and annihilates them by a Jacobi rotation. Successive rotations undo previously set zeros, but the norm of the off-diagonal elements

$$\text{off}(M) = \sqrt{\sum_{i \neq j} |M_{ij}|^2} \qquad (25)$$

nevertheless decreases monotonically until the matrix is diagonal to machine precision. The matrix of eigenvectors is obtained almost as a by-product, simply by accumulating the product of the Jacobi rotations. While in the literature [16,1,12] the method is discussed for real symmetric matrices, the generalisation for complex hermitian matrices is straightforward if one uses the parametrisation of the rotation matrices given in [17]; one simply has to keep trace of the additional complex phase of the off-diagonal elements. The convergence we found in our tests is very fast, typically within a few sweeps, and thus the CPU cost is negligible in comparison with the CG part.

In order to obtain the eigenvalues of $M$ in increasing order, $\nu_1 \leq \ldots \leq \nu_n$, the Jacobi diagonalizer has to be combined with a routine which sorts the eigenvalues and rearranges the eigenvectors of $M$ correspondingly. The need for this sorting arises from the assumption that $w'_i$ is a better approximation to a low-lying eigenvector than $w'_j$ if $\nu_i \equiv \mu(w'_i) < \mu(w'_j) \equiv \nu_j$. Moreover, it may happen that the (pure) CG part on its own produces approximations $w_i$ which do not satisfy $\mu(w_1) \leq \ldots \leq \mu(w_k)$. This kind of "level crossing" occurs whenever the search for some higher eigenvalue finds a component proportional to a low-lying eigenvector which is not yet in the span of the previously found approximate eigenvectors. The sorting is also necessary for using the Temple bound as a stopping criterion.

## 3.3 Length of the CG cycles

In step (i) of our algorithm (see sec. 2.2) the CG search for the $k$-th eigenvalue is terminated after at most $N(k)$ iterations (or before, as soon as the precision required by the stopping criterion has been reached). So far, we have not yet specified how this maximal length of the CG searches should be chosen.[11] The simplest procedure would be to take a fixed value $N(k) = N_0$ for all CG search cycles and for all eigenvalues. The optimal choice of $N_0$ will then depend on the properties of the matrix $A$, in particular on those which determine the rate of convergence of the algorithm, like the dimension of $A$ or/and the (unknown) spectrum itself.

Looking for a strategy to find an optimal choice of $N_0$ we have investigated the number of iterations which are spent for the individual eigenvalues. It turns out that the lower eigenvalues generally need least iterations for relatively small values of $N_0$. This is to be expected because the lower eigenvalues gain from frequent intermediate diagonalizations. On the other hand, the higher eigenvalues prefer larger values of $N_0$. This is plausible for two reasons: First, the fewer iterations have been spent on the lower eigenvalues, the worse is the approximation of the projectors $P_{k-1}^\perp$ by the $Q_{k-1}^\perp$. The purpose of these projections is to keep the search directions for the $k$-th eigenvalue orthogonal to the space spanned by the eigenvectors belonging to eigenvalues $\lambda_l$ with $l < k$. Therefore, a large misorientation between $P_{k-1}^\perp$ and $Q_{k-1}^\perp$ may lead to CG steps with larger components in the unwanted directions of the lowest eigenvectors. Second, for the higher eigenvalues, which in general have a worse rate of convergence, the loss due to restarting the CG search seems to be larger than for the lower eigenvalues. Finding the optimal value of $N_0$ corresponds therefore to a compromise between lower eigenvalues, which converge fastest for small $N_0$, and higher eigenvalues, which need less iterations for large $N_0$.

One may try to reduce the loss from restarting the CG algorithm by storing for each eigenvalue the last search direction $p_k$ of a CG cycle. Then, after performing the intermediate diagonalization step, one restarts the next CG cycle for the $k$-th eigenvalue with initial search directions, which are obtained by rotation analogous to (19):

$$p'_k = \sum_{i=1}^{n} \xi_i^{(k)} p_i \ . \tag{26}$$

Indeed, if the algorithm is likewise modified the best performance (i.e. the least total number of CG iterations) is achieved for smaller values of $N_0$ than without saving the search directions. Moreover, the same value of $N_0$, which is optimal for the total number of iterations, then turns out to be also optimal for most of the individual eigenvalues. However, for the cases we tested, the total number of iterations needed for the modified algorithm was not significantly smaller than in the original version without saving the search directions and, of course, the value of $N_0$ is still a parameter which needs to be tuned.

---

[11] For the sake of notational simplicity we do not explicitly indicate the fact that $N(k)$ may be different in each CG cycle.

To avoid the problem of tuning the value of $N_0$ we propose a different strategy in which $N(k)$ is not fixed but rather determined independently for each CG cycle and for each eigenvalue by a suitable criterion. Motivated by the role played by an adequate precision of the projectors $Q_{k-1}^\perp$, we determine the length of the CG search cycles by the requirement that the error of the Ritz functional for the current eigenvalue has been decreased at least by a factor of $\gamma^{-1} = O(10)$. Using the convergence proportional to $\|g\|^2$, this amounts to running the CG searches for each eigenvalue until

$$\frac{\|g_i\|^2}{\|g_0\|^2} \leq \gamma \,, \tag{27}$$

where $g_0$ and $g_i$ are the gradient of the corresponding Ritz functional in the first and last iteration of the present CG search, respectively. Proceeding according to (27) ensures a simultaneous and homogeneous convergence for all eigenvalues. The CG searches for higher eigenvalues automatically become longer because of their slower convergence. Similarly the lengths of the CG searches are automatically adjusted appropriately in the different cycles during the algorithm.

The choice of the ratio $\gamma$ is rather uncritical and in practice does not need to be tuned even when treating quite different matrices $A$. We found $\gamma \sim 0.1$ to be a good value in all our tests, and the corresponding total number of iterations needed for convergence was comparable or often less than what could be achieved with the optimal choice for a fixed $N_0$. Changing the ratio $\gamma$ to 0.05 or 0.2, the total number of required iteration varies by at most 10%, and in most cases it is increased.

In addition to the criterion for the reduction of $\|g\|^2$ according to (27), it is advisable to impose a minimal and maximal value on $N(k)$. Restricting the lowest value of $N(k)$ to at least a few, say 5, iterations prevents the first few CG cycles, when the decrease of $\|g\|$ can be very fast, from being unreasonably small. The maximal search length should be very large, $O(100)$ or bigger, and this cut-off should become effective only in rare cases when wasting too much iterations in an extremely slow convergent CG search (which may benefit from the intermediate diagonalization).

If different matrices $A$ are treated at the same time on a SIMD parallel com-puter, it is possible to require that condition (27) is fulfilled at least on one node or on all nodes, which means that $N(k)$ is determined by the fastest system or by the slowest one, respectively. If the spectrum for the different matrices is comparable, both implementations are possible and either of them can be faster by up to 20%. Nevertheless, we prefer the first variant in order to insure a minimal decrease of $\|g\|$ for all systems.

## 3.4 Stopping criteria

We based the stopping criterion on three different estimates of the (relative) error of the Ritz functional. In the numerical tests, we monitored the real error by comparing with reference eigenvalues computed by a Lanczos algorithm [6].

The first stopping criterion exploits the rigorous error estimate proportional to $\|g\|$. Before reaching the asymptotic convergence of the CG searches, i.e. when evaluating the eigenvalues only with a crude relative error in the per-mille range, this estimate is not yet too pessimistic and the other estimates to be described below are not yet reliable enough. When running the algorithm for a higher precision of the eigenvalues, the error estimate based on $\|g\|$ will soon become too pessimistic by several orders of magnitude, because it does not take into account the convergence quadratic in $\|g\|$.

A more realistic error estimate, and hence a more efficient stopping criterion, is obtained by Temple's inequality. To apply it to the $k$-th eigenvalue a lower bound $l_{>k}$ for the next-to-$k$th lowest eigenvalue $\lambda_{>k}$ is needed. Since these quantities are unknown a priori, one approximates $l_{>k}$ by the value of the next Ritz functional $\mu(w_j)$, $j \geq k+1$, which is bigger than $\mu(w_k)$ (to the level of the required precision). Although this choice does not satisfy $\lambda_{>k} \geq l_{>k}$, as required for (22), it turns out that the resulting error estimate usually remains very safe (see below and also ref. [19]).

In practice, we do not use the Temple criterion during the first few CG cycles and as long as we do not have $\delta_T(x_0) < \|g_0\|$ at the beginning of a new CG cycle. This condition insures [12] that the approximation of $l_{>k}$ by $\mu(w_{>k})$ in the denominator of $\delta_T$ is not too bad. As long as $\delta_T(x_0) < \|g_0\|$ is not satisfied, one finds that $\delta_T$ does not evolve smoothly, but "jumps" to higher values at the beginning of the CG cycles when the estimate for $l_{>k}$ is significantly decreased by the intermediate diagonalization.

From the comparison with the Lanczos data we know that the proportionality factor in front of $\|g\|^2$ in the Temple bound (24) is often too conservative by up to a factor of order 100. For this reason we have implemented a third stopping criterion based on the comparison of the values of the Ritz functional $\mu$ and $\mu'$ after two subsequent intermediate diagonalizations. We assume that (at least after the startup phase when geometric progression is reached) the actual error is reduced during a CG cycle by about the large factor $\gamma^{-1}$ of (27) by which $\|g\|^2$ has decreased. This yields an (a posteriori) estimate for the error at the beginning of the latest complete cycle,

$$\delta_{\text{cycle}} = \frac{\mu - \mu'}{1 - \gamma} \ . \tag{28}$$

---

[12] $\delta_T(w_k) < \|g(w_k)\|$ implies that the denominator of $\delta_T$ is strictly (in practice several orders of magnitude) bigger than $|\lambda_k - \mu(w_k)|$, cf. (18). The latter is of the same order as the effect of using $\mu(w_{>k})$ instead of $\lambda_{>k}$ for $l_{>k}$.

The decrease of the error during this last completed CG cycle itself might be estimated by an additional factor of $\gamma$. However, to remain on the safe side, we refrain from this and use $\delta_{\text{cycle}}$ as the initial error estimate at the beginning of only the following CG cycle. In order to have an actual stopping criterion also in each CG iteration during the following (incomplete) CG cycle we then extrapolate $\delta_{\text{cycle}}$ according to the decrease in $\|g\|^2$. After the next intermediate diagonalization $\delta_{\text{cycle}}$ is again updated according to (28).

This stopping criterion based on $\delta_{\text{cycle}}$ saves typically another $20 - 40\%$ of the total number of iterations compared to using (only) the Temple criterion. In particular, for the highest (and usually most expensive) eigenvalue the Temple criterion cannot be used anyhow, such that one has at hand only the inefficient gradient criterion otherwise.

In fig. 1 we show the convergence for the lowest (a) and the 14th-lowest (b) eigenvalue together with the various error estimates for one of our test configurations. Note the dip in the curve for the true error $\mu - \lambda_{14}$ in fig. 1(b). This is due to picking up some component from eigenvectors with $\lambda < \lambda_{14}$ in course of the CG search. In such situations the intermediate diagonalizations stabilize the convergence and undo the "level crossing".

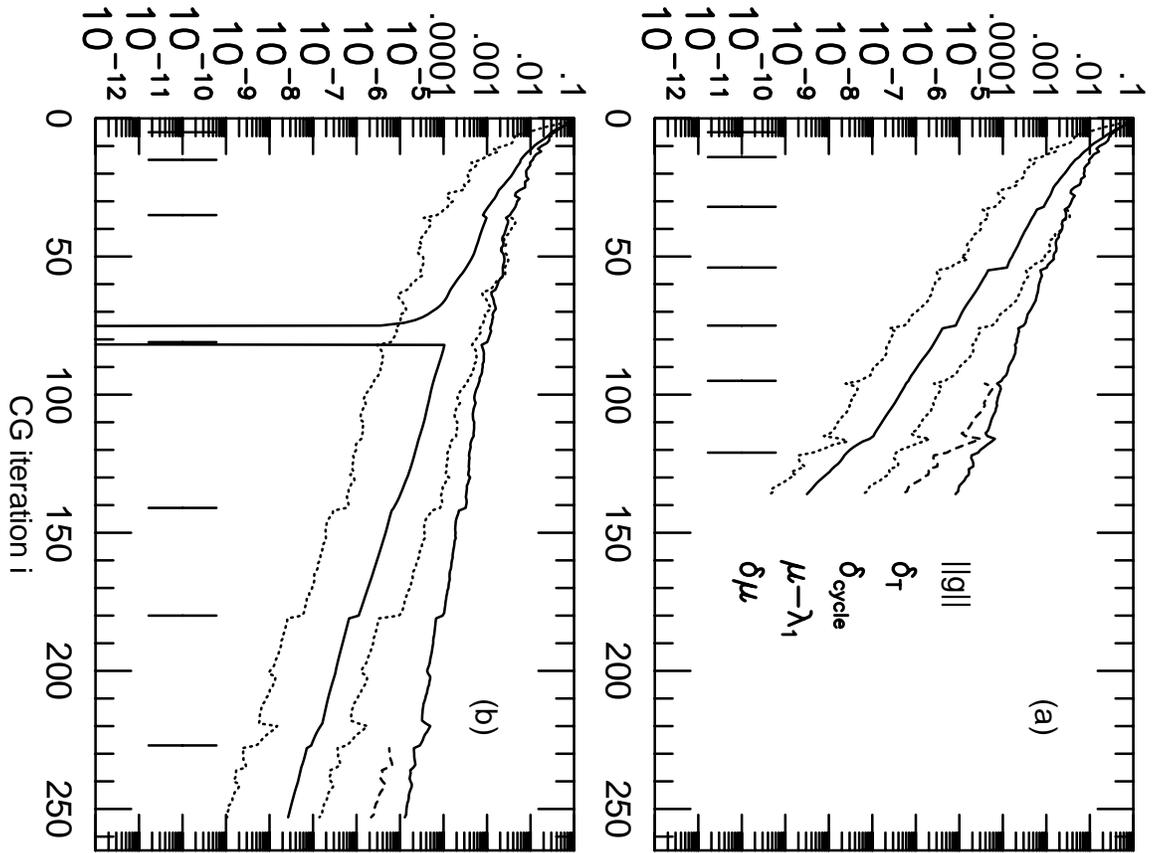

Fig. 1. Convergence of the Ritz functional for $\lambda_1$ (a) and $\lambda_{14}$ (b) of the operator $\mathcal{Q}^2$ (cf. (30)) on the $8^3 \cdot 12$ lattice (cf. table 1). Shown is the true error $\mu - \lambda$, the various error estimates discussed in the text, and the decrease $\delta\mu$ of the Ritz functional per CG iteration. The vertical bars indicate where the intermediate diagonalizations took place. The curves in (b) show the analogous quantities as labeled in (a).

## 4 Performance tests

### 4.1 The Wilson-Dirac operator in lattice QCD

We tested the accelerated CG method of sec. 2 for the case of the lattice Dirac operator $(D+m)$ with Wilson fermions of bare mass $m$ in SU(2) gauge fields with periodic boundary conditions. On a four-dimensional lattice of sites $x$,

$(D+m)$ acts on a lattice spinor $\psi$ as follows,[13] see e.g. [20],

$$[(D+m)\psi]^{\alpha a}(x) =$$
$$\frac{1}{2\kappa}\psi^{\alpha a}(x) - \frac{1}{2}\sum_{\mu=1}^{4} \{ (\mathbb{1} - \gamma_\mu)_{\alpha\beta} U(x, x+\mu)_{ab} \psi^{\beta b}(x+\mu)$$
$$+ (\mathbb{1} + \gamma_\mu)_{\alpha\beta} U(x, x-\mu)_{ab} \psi^{\beta b}(x-\mu)\} \ . \quad (29)$$

Here $\kappa = (2m+8)^{-1}$ denotes the hopping parameter and $x \pm \mu$ is the nearest neighbour site of $x$ in $\pm\mu$-direction. The gauge field $U(x, x \pm \mu) \in \mathrm{SU}(2)$ lives on the links $(x, x \pm \mu)$ of the lattice. $U$ is generated by some Monte-Carlo process, see e.g. [20]. On the rhs of eq. (29) an implicit summation over the spinor indices ($\beta = 1, \ldots, 4$) and colour ($b = 1, 2$) is understood.

The operator which we consider is $A = \mathcal{Q}^2$ with

$$\mathcal{Q} = \gamma_5 (D+m) / (8+m) \ . \quad (30)$$

Thus $A$ is hermitian and it is normalised such that its eigenvalues are between 0 and 1. Note that all steps of the CG minimisation and of the Jacobi diagonalization are gauge covariant; hence, no gauge fixing is required. For typical gauge fields $U$ the distribution of the eigenvalues of the operator $\mathcal{Q}^2$ is relatively smooth without exceptional gaps [6]. Our tests with $A = \mathcal{Q}^2$ are special, but we expect the numerical results to be comparable for other operators which have a spectrum similar to $\mathcal{Q}^2$.

### 4.2 Numerical results

In table 1 we give an overview of some of our numerical tests. All gauge fields at finite $\beta$ (the coupling constant of the gauge part of the action, see e.g. [20]) were generated in the presence of two flavours of dynamical fermions. In the table we also give the lowest eigenvalue and the average gap $\langle\Delta\rangle_{16}$ among the lowest 16 eigenvalues. We required a relative accuracy of $10^{-4}$ according to any of the three stopping criteria discussed in sec. 3.4; in practice $\delta_{\mathrm{cycle}}$ is most efficient and hence gets relevant in most instances. On APE/Quadrics computers the code runs with an efficiency of above 35% of the peak performance. "Q1sec" and "QH2sec" in the last row of the table refer to the actual time on a 8 and 256 node machine, respectively.

---

[13] The hermitian Euclidean $\gamma$-matrices $\gamma_\mu$, $\mu = 1, 2, 3, 4$, satisfy the Clifford algebra $\{\gamma_\mu, \gamma_\nu\} = 2\delta_{\mu\nu}\mathbb{1}$. Moreover, $\gamma_5 \equiv \gamma_1\gamma_2\gamma_3\gamma_4$ anticommutes with all of them and $\gamma_5^2 = \mathbb{1}$ is the $4 \times 4$ unit matrix.

Table 1
Some of the lattice sizes and gauge configurations used in the tests. The upper part refers to physical properties and the lower part summarizes the performance of the algorithm. The last row refers to the actual computer time required for 8 eigenvalues (cf. text).

| lattice size | $4^4$ | $8^4$ | $8^3 \cdot 12$ | $16^4$ |
|---|---|---|---|---|
| $\beta$ | 1.75 | 0.00 | 2.12 | 2.12 |
| $\kappa$ | 0.15 | 0.20 | 0.15 | 0.15 |
| $\lambda_1$ | $6.513 \cdot 10^{-3}$ | $1.592 \cdot 10^{-3}$ | $8.098 \cdot 10^{-4}$ | $7.703 \cdot 10^{-4}$ |
| $\langle \triangle \rangle_{16}$ | $\approx 5 \cdot 10^{-4}$ | $\approx 1 \cdot 10^{-5}$ | $\approx 7 \cdot 10^{-5}$ | $\approx 1 \cdot 10^{-5}$ |
| # eigenvalues | # iterations for a relative accuracy of $10^{-4}$ | | | |
| 8 | 690 | 4120 | 2130 | 4340 |
| 16 | 1260 | 5730 | 3540 | 7070 |
| 32 | 2040 | 10480 | 4810 | 13780 |
| 64 | 3110 | 15950 | 8640 | 19960 |
| 8 | 9 Q1sec | 405 Q1sec | 345 Q1sec | 225 QH2sec |

In addition to the lattices quoted in table 1, we studied random configurations ($\beta = 0$) on various lattice sizes with $\kappa \geq 0.2$, which is relatively close to the critical value where (almost-)zero modes arise. In order to verify that degeneracies are obtained correctly by the algorithm, we have studied trivial gauge configurations $U \equiv \mathbb{1}$ corresponding to free quarks or $\beta = \infty$. In this case every eigenvalue is at least eightfold degenerate, because the free $\mathcal{Q}^2$ is diagonal in Dirac and colour indices. The parallelization of the algorithm when treating independent systems on different nodes has been investigated using eight independent unquenched $6^3 \cdot 12$ lattices at $\beta = 2.12$, $\kappa = 0.15$. In all cases we found satisfactory performances.

A typical plot of the convergence of the Ritz functional and of the various error estimates is shown in fig. 1 for the $8^3 \cdot 12$ lattice. Qualitatively the same behaviour is found for the other lattices. The differences occur in the rate of convergence.

For the CG minimisation of quadratic forms the asymptotic rate of convergence is determined by the square root of the condition number of the Hessian matrix. Similarly one expects that the asymptotic convergence for the pure CG minimisation of the Ritz functional for the $k$-th eigenvalue is governed by the ratio

$$c_k = \frac{\lambda_{k+1} - \lambda_k}{\lambda_{\max} - \lambda_k} \; , \tag{31}$$

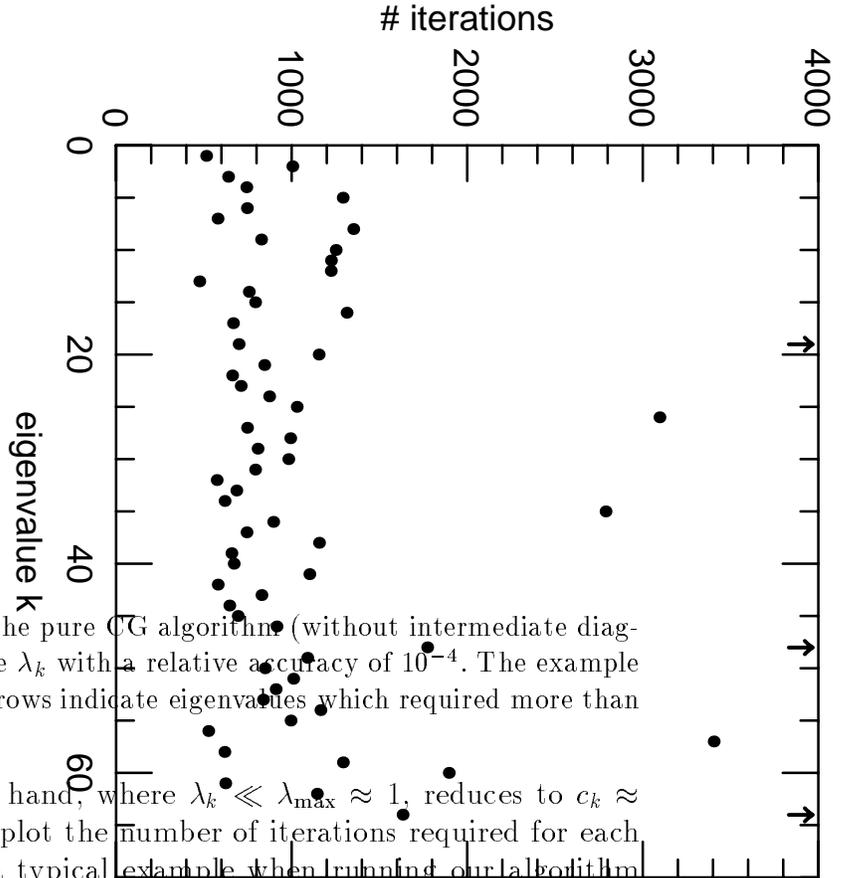

Fig. 2. Number of iterations in the pure CG algorithm (without intermediate diagonalization) required to compute $\lambda_k$ with a relative accuracy of $10^{-4}$. The example is taken on the $8^3 \cdot 12$ lattice. Arrows indicate eigenvalues which required more than 4000 iterations.

which in the case $A = \mathcal{Q}^2$ at hand, where $\lambda_k \ll \lambda_{\max} \approx 1$, reduces to $c_k \approx \lambda_{k+1} - \lambda_k \equiv \triangle_k$. In fig. 2 we plot the number of iterations required for each of the lowest eigenvalues in a typical example when running our algorithm without intermediate diagonalizations (i.e. with $\gamma = 0$ or $N(k) = \infty$). The number of iterations strongly fluctuates and is indeed closely correlated with the inverse square root of the gap to the next eigenvalue.

On the other hand, when the length of the CG cycles is chosen according to (27) and the intermediate diagonalizations are performed, the convergence becomes quite different as shown in fig. 3. Most obvious is the fact that the numbers of iterations required for the individual eigenvalues lie on relatively smooth curves $f(k, n)$ where $n$ denotes the total number of eigenvalues which were determined. Generally one observes that the eigenvalues with index $k$ near $n$ take longest to converge, and that the number of iterations decreases rapidly for eigenvalues with smaller index. When running with different $n$ the behaviour of $f(k, n)$ is rather similar and seems to depend essentially only on the ratio $k/n$, i.e. on the relative position among the eigenvalues which are calculated. The total number of iterations required for a given number of eigenvalues roughly follows a linear increase with $n$, at least for moderately

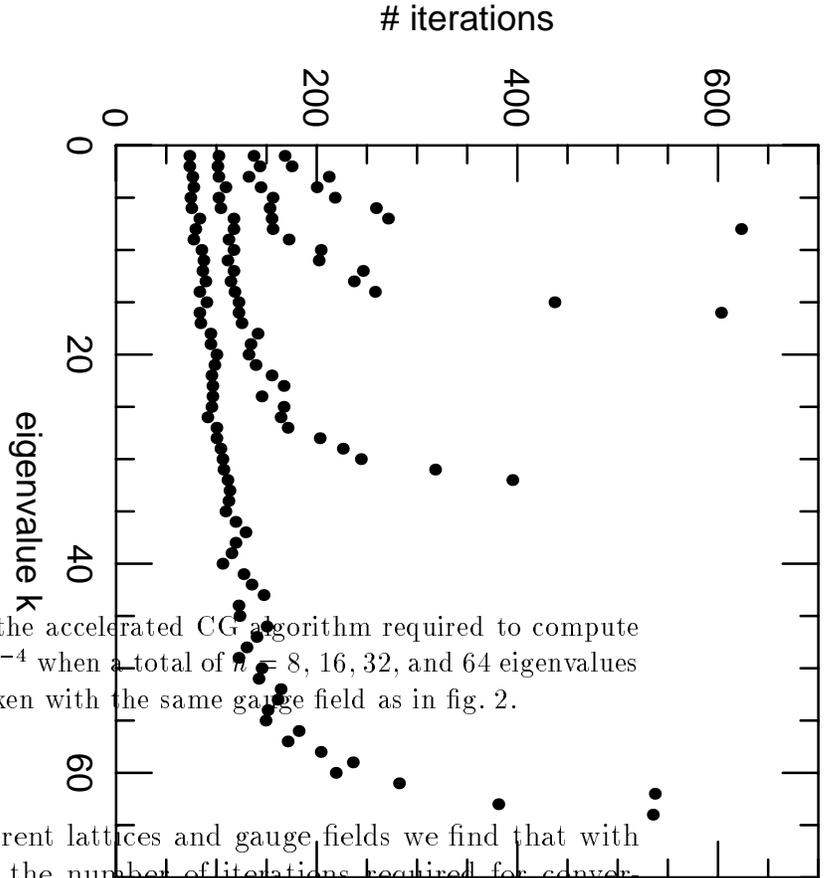

Fig. 3. Number of iterations in the accelerated CG algorithm required to compute $\lambda_k$ with a relative accuracy of $10^{-4}$ when a total of $n = 8, 16, 32,$ and $64$ eigenvalues is computed. The example is taken with the same gauge field as in fig. 2.

small $n$ up to order 100.

Comparing the results of different lattices and gauge fields we find that with intermediate diagonalizations the number of iterations required for convergence is governed by the average $\langle \triangle \rangle_n$ for the first $n$ gaps. Empirically, the number of iterations $N_{it}(k)$ needed for the $k$-th eigenvalue to converge can thus be described approximately (within about a factor of two) by

$$N_{it}(k) \approx \sqrt{\frac{\lambda_{\max} - \lambda_1}{\langle \triangle \rangle_n}} f\left(\frac{k}{n}\right) , \qquad (32)$$

where $f$ depends somewhat on the overall properties of the spectrum, but is almost identical for configurations with the same physical parameteres $\beta$ and $\kappa$.

In terms of the total number of CG iterations for all eigenvalues we find typically a gain of a factor of $4 - 8$ compared to running a pure CG algorithm without intermediate diagonalizations (or restarts, which by themselves would make the convergence only worse).

Of course, the pure number of CG iterations is not the only factor which determines the total computer time. The work required for the projections by $Q_k^\perp$ grows linearly with $k$ and therefore adds a component to the total computer time which grows quadratically with $n$. If the state renormalizations and projections (see also the comment at the end of sec. 3.1) are done every 10-th iteration, the average time spent for applying $Q_{k-1}^\perp$ is about $(k-1)\times$ 4% of the time of one CG iteration for the lowest eigenvalue. Due to the fact that the higher eigenvalues ($k \lesssim n$) are more expensive in terms of computer time per iteration and that they take most iterations to converge, the gain in computer time may be somewhat less than in the number of ($\mathcal{Q}^2 \times$vector) multiplications.

The remarkable uniformity of the convergence, i.e. the fact that $f(k,n)$ is a relatively smooth function of $k$ (fig. 3) compared to the individual numbers of iterations needed without intermediate diagonalizations (fig. 2), is a clear benefit for SIMD parallelisation when several systems are treated simultaneously: It reduces the idle time while some of the processors have to wait until the corresponding eigenvalue has converged for all systems.

As to the effect of introducing "dummy" eigenvalues, i.e. of increasing $n$ by a few eigenvalues without requiring the stopping criterion for them, it depends on the size of the gap $\triangle_n$ for the last eigenvalue[14]. We observed that an overall gain in the net number of iterations is only achieved if the gaps for the additional dummy eigenvalues are comparable or larger than the gaps for the last (few) eigenvalues which are treated fully. Usually there is no advantage from introducing more than $5 - 10\%$ of $n$ as dummy eigenvalue (or at least one for small $n$), but no general rule can be recommended. On the other hand, if the information about the less precise dummy eigenvalues is of interest in a particular application, their possible cost may be worthwhile in any case.

## 5   Conclusions

We presented an accelerated CG algorithm for the computation of the low-lying eigenvalues of a hermitian operator $A$. This algorithm was tested for the case of the squared Dirac operator $A = \mathcal{Q}^2$ in lattice gauge theory. The key features of our algorithm are the following.

– Rigorous error bounds can be derived just from the last CG iterate.
– The correct multiplicities are detected.

---

[14] It also depends on whether $\delta_{\text{cycle}}$ is used as a stoppng criterion, or whether one uses only the Temple estimate, which is not applicable for the last eigenvalue (or for more than just the last eigenvalue in case that degeneracies are present).

– Approximations to eigenvectors are obtained as a by-product.
– The pure CG algorithm is speeded up through the intermediate diagonalizations by a factor of $4-8$.

Comparing only the performance, the accelerated CG algorithm is still inferior to the Lanczos method. With the studied configurations and numbers of eigenvalues, the accelerated CG algorithm needs about $5-8$ times more ($\mathcal{Q}^2\times$vector) multiplications than a Lanczos method for the (unsquared) operator $\mathcal{Q}$ [6]. It is hard to say how this factor of $5-8$ converts to CPU time, because the two algorithms were implemented on different computers, and one also has to consider the work that has to be done apart from the ($\mathcal{Q}^2\times$vector) or ($\mathcal{Q}\times$vector) operations, respectively.

A time-consuming part in the CG algorithm are the repeated projections onto the subspaces orthogonal to previously computed approximate eigenvectors. However, in practice it is not necessary to perform these projections in every iteration. This shows that the algorithm is numerically very stable. The accelerated CG algorithm can be implemented on (SIMD) parallel computers such that even different matrices can be treated simultaneously in an efficient way.

Compared to a Lanczos method without any re-orthogonalisation [2] we need more computer memory in order to store the approximate eigenvectors. However, in view of today's computer capabilities we do not consider this as a real disadvantage. Also, in certain applications one needs the eigenvectors when one is interested in the contribution of the low-lying eigenmodes to physical observables. Computing the eigenvectors in a separate step (for instance by inverse iteration) after the determination of the eigenvalues would be much more expensive.

Despite a superiority of the Lanczos procedure when viewed from the CPU time point-of-view (even if only a few eigenvalues are required), we consider the algorithm presented here favorable. In particular, in the Lanczos algorithm one does not have any rigorous estimate of the numerical accuracy and convergence can only be estimated from experience [6]. Moreover, a Lanczos method does not yield any information about degeneracies in the spectrum.

**Acknowledgement**

The idea to accelerate the CG algorithm of ref. [5] by intermediate diagonalizations and valuable suggestions for the stopping criterion are due to M. Lüscher, and we essentially used his TAO code for the CG part of the algorithm. We wish to thank B. Bunk, M. Lüscher, and U. Wolff for helpful discussions, and C. Liu and K. Jansen for providing us with the unquenched gauge configura-tions. TK gratefully acknowledges financial support by Deutsche Forschungsgemeinschaft under grant Wo 389/3-1. The computations reported here were performed on the APE/Quadrics machines of DESY and on workstations of the Humboldt University.